\documentstyle[graphicx]{article} 

\date{~}

\begin{document}

\title{%
\vskip-1.0in\vskip-6pt \hfill {\rm\normalsize UCLA/02/TEP/22} \\ 
\vskip-12pt \hfill {\rm\normalsize CWRU-P11-02} \\ 
\vskip-12pt \hfill {\rm\normalsize NSF-ITP-02-96} \\ 
\vskip-0pt~\\
Indirect detection of a subdominant density component of cold 
dark matter}

\author{Gintaras Duda, Graciela Gelmini \\ 
  {\it Department of Physics and Astronomy, UCLA} \\ 
  {\it Los Angeles, CA 90095-1547, USA} \\
  {\tt gkduda,gelmini@physics.ucla.edu} \\ 
  \and 
  Paolo Gondolo \\ 
  {\it Department of Physics, Case Western Reserve University} \\
  {\it 10900 Euclid Ave., Cleveland, OH 44106-7079, USA} \\
  {\tt pxg26@po.cwru.edu} 
  \and 
  Joakim Edsj\"o \\ 
  {\it Department of Physics, Stockholm University} \\ 
  {\it AlbaNova, SE-106 91, Stockholm, Sweden} \\ 
  {\tt edsjo@physto.se} 
  \and 
  Joseph Silk \\ 
  {\it Department of Physics, University of Oxford} \\ 
  {\it Denys Wilkinson Building, Keble Road, Oxford OX1 3RH, UK} \\
  {\tt silk@astro.ox.ac.uk} }

\maketitle

\begin{abstract}
  \vbox{We examine the detectability through indirect means of weakly
    interacting dark matter candidates that may constitute not all but only a
    subdominant component of galactic cold dark matter.  We show that the
    possibility of indirect detection of neutralinos from their annihilations
    in the Earth and Sun is not severely hampered by decreasing neutralino
    relic density.  Upward-going muon fluxes in underground detectors from
    neutralino annihilations in the Sun can remain above the threshold of
    detectability of 10 muons/km$^2$/yr for neutralinos composing 1$\%$ or more
    of the halo dark matter.  Similarly, signals from neutralino annihilations
    in the Earth can also remain high for neutralino densities of 1$\%$ of the halo
    and actually would only be observable close to this low density for
    neutralinos lighter than 150 GeV.  We also show that there are many models
    which simultaneously have high direct and indirect detection rates making
    some model discrimination possible if a signal is seen in any of the
    current dark matter searches.  }
\end{abstract}

\newpage

We examine the observability of collisionless cold dark matter (CCDM) by
indirect detection when it is merely a subdominant component of the cold dark
matter.  Originally we were led to consider such a question by claims of the
need for collisional cold dark matter \cite{steinhardt} as the main form of
dark matter in the universe, but the question is valid independently of such
claims.  Even if non-CCDM is proven to be unnecessary, there remains the
possibility of the CDM consisting of several populations, the one we are
searching for not being the dominant one. So the question is: if the previously
favored CDM candidates, such as Weakly Interacting Massive Particles (WIMPs),
constitute only a fraction, say 1\% or less, of the local dark matter halo,
would these particles still be observable in the current and proposed direct
and indirect dark matter searches?  We could even reverse our question in the
following manner: if we see a CDM signal in any of our searches, could we be
observing a subdominant component of the total CDM?  Additionally, could a
combination of direct and indirect detection data resolve the issue of which
component of the total CDM we are observing?

Unless there is segregation for different types of cold dark matter, the ratio
of collisionless CDM to total CDM should be the same locally in the Galaxy and globally in
the whole Universe. This is expected if all CDM components are collisionless.
Thus in the following we assume that the local fraction of the particular
collisionless CDM $f_{CCDM}$ is related to the CCDM density parameter $\Omega_{CCDM}$ (the
CCDM relic density in units of the critical density) through
\begin{equation} \label{no-segregation} f_{CCDM} =
  \frac{\rho_{CCDM}}{\rho_{\rm local}} = \frac{\Omega_{CCDM}}{\Omega_{CDM}}.
\end{equation}
Here $\rho_{CCDM}$ is the local density of a particular CCDM candidate, such as
a particular WIMP, $\rho_{\rm local}\simeq$ 0.3 GeV/cm$^3$ is the local halo
density (at the location of the Earth), $\Omega_{CCDM}$ is the relic density of
our particular CCDM candidate, and $\Omega_{CDM} \simeq$ 0.3 is the total
contribution of DM to the total energy density of the Universe.

Naively, one may expect that if the local CCDM density is, say, $f_{CCDM} =
1\%$ of the local halo density, the fluxes of particles produced by WIMP-WIMP
annihilations in the halo and elsewhere, being proportional to some power $n$
of the local WIMP density, would decrease by an amount $f_{CCDM}^n$, i.e.\ by
$0.01^n$ in our example.  However, the relic density of thermal WIMPs
$\Omega_\chi$ is approximately determined by their annihilation cross-section $\sigma_a$
through the relation
\begin{equation}
  \Omega_\chi h^2 \simeq \frac{3 \times 10^{-27} 
    {\rm cm^3/sec}}{\left<\sigma_a v\right>_{E.U.}},
  \label{eq:1}
\end{equation} 
where $\left<\sigma_a v \right>_{E.U.}$ is the thermal average of the
annihilation cross section times the relative velocity of the WIMPs at
freeze-out, in the early universe, and $h$ is the reduced Hubble constant,
$h\simeq 0.7$.  Hence a reduction in the relic WIMP density requires an
increase in their annihilation cross-section in the early Universe.  In
general, $\left<\sigma_a v \right>_{E.U.}$ is related to $\left<\sigma_a
  v\right>$, the annihilation cross section of WIMPs in the galactic halo or at
the center of the Sun or of the Earth. These two averaged annihilation cross
sections differ in the temperature of the WIMPs when they annihilate:
about a twentieth of the WIMP mass in the early universe, practically zero in
the halo, the Sun and the Earth. However, there are exceptions to this
relation: when coannihilations \cite{coann} are important the WIMP population has to be
followed together with the population of other particles with which it can be
interchanged. As a consequence, $\left<\sigma_a v \right>_{E.U.}$ is an
effective annihilation quantity that includes annihilation of and with other
particles, and which is not directly related to $\left<\sigma_a
  v\right>$ at zero temperature.
Since the annihilation signals are in general
proportional to the product of the annihilation cross section and the {\it
  square} of the WIMP density, and the latter decreases linearly with the
annihilation cross section, a more educated statement is that the annihilation
signals decrease linearly with WIMP density. This is the case if
$\left<\sigma_a v\right>_{E.U.}$ and $\left<\sigma_a v\right>$ are
proportional.  We will see below (Fig. 3) that this linear decrease is true for
the envelope of the largest signals of neutralino annihilation in the halo.

The situation is more complicated when the signals depend not only on the
annihilation cross section $\sigma_a$, but also on the scattering cross section
$\sigma_s$ of WIMPs with nucleons. Besides direct detection, this is important
for WIMP annihilation in the Sun and the Earth. The number of WIMPs accumulated
in those bodies is proportional to $\sigma_s \rho_{CCDM} $, i.e. to $\sigma_s
\sigma_a^{-1}$. So at short times the annihilation signal goes as $\sigma_a
(\sigma_s \sigma_a^{-1})^2$.  At times much longer than the equilibration time
between WIMP capture and annihilation, instead, the annihilation signal depends
entirely on the product $\sigma_s \rho_{CCDM} \sim \sigma_s \sigma_a^{-1}$,
since in equilibrium the annihilation rate equals the capture rate.

Now, the scattering cross section is not independent of the annihilation cross
section.  As argued in earlier papers \cite{MANY, dgg, bottino2341}, when the
annihilation cross section grows, the scattering cross section may grow by the
same factor.  If so, the product $ \sigma_s \rho_{CCDM} \sim \sigma_s
\sigma_a^{-1} $ remains approximately constant, and signals proportional to
this product, such as those from equilibrium annihilation in the Sun and Earth,
remain nearly unchanged.

Scattering and annihilation cross sections of WIMPs are in general related by a
crossing symmetry. Unpolarized cross sections are spin sums of amplitudes
squared, and a crossing symmetry relates the amplitudes of two processes that
differ by having a particle in the final state exchanged with a particle in the
initial state, both becoming their respective antiparticle.  For example, the
amplitude for neutralino-quark scattering $\chi q \to \chi q$ is related to the
amplitude for neutralino-neutralino annihilation into quark-antiquark pairs
$\chi\chi\to q \overline{q}$. Often, other annihilation channels besides
$q\overline{q}$ are open, such as annihilation into W or Z boson pairs, and
since these particles are not present in the nucleon, the crossing symmetry is
imperfect. Moreover, the scattering and annihilation cross sections are
computed at different kinematical points, and may differ in the spin structure,
the presence of resonances, the integration limits over momenta, etc.

The relation between the WIMP annihilation cross section and the cross section
for WIMP scattering off nucleons is further complicated by the fact that
annihilation can be considered to produce quark-antiquark pairs, while
scattering occurs with quarks bound in nucleons.  The annihilation amplitudes
are functions of the quark masses, while the scattering amplitudes are
functions of the nucleon mass.  Moreover, the scattering amplitude contains the
matrix elements of the quark currents in the nucleon, which are of course not
present in the annihilation amplitude into quark-antiquark pairs.

Despite these imperfections in the crossing symmetry, we can see it at work in
the case where WIMPs are neutralinos in the Minimal Supersymmetric Standard
Model (MSSM), with some sets of parameters. We use the version of the MSSM
described in \cite{bergstrom-gondolo}, whose seven input parameters are given
at the weak scale, as implemented in the DarkSUSY code~\cite{darksusy}. In
Fig.~1 we make the crossing symmetry explicit in a particular example. 

We
present a set of models obtained by modifying only the neutralino composition
(and thus its couplings) starting from a single original model that exhibits
high muon fluxes from neutralino annihilations in the Earth at low neutralino
relic density (we argue later that this high flux is due directly to the
compensation provided by the crossing symmetry).  The particular parameter
values of the original model (one of the points in Fig.~1) are as follows: the
mass of the neutralino $m_\chi$ = 147.9 GeV, the mass of the CP-odd Higgs boson
$m_A$ = 113.1 GeV, the ratio of Higgs vacuum expectation values $\tan\beta$ =
47.52, the sfermion mass parameter $m_0$ = 1873 GeV, and the trilinear
couplings in the top and bottom sectors $A_t$ = -1.758$m_0$ and $A_b$ =
-2.152$m_0$.  This particular original model has a neutralino relic density of
$\Omega_\chi h^2 = 2.9 \times 10^{-3}$, a neutralino gaugino fraction $Z_g$ =
0.2511, and muon fluxes from neutralino annihilations in the Sun and Earth of
$\Phi_\mu^\odot$ = 143.3 muons/km$^2$/yr and $\Phi_\mu^\oplus$ = 44.25
muons/km$^2$/yr, respectively.  In order to examine how the scattering cross
section changes with the annihilation cross section, we varied the higgsino and
gaugino mass parameters of the original model, $\mu$ and $M_2$, while keeping
the neutralino mass and all the other MSSM parameters fixed.  This variation
produces a ``line'' in the model parameter space with constant neutralino mass
but varying $\mu$ and $M_2$ parameters, and hence different gaugino and
higgsino compositions. Fig.~1 shows plots of the spin-independent
neutralino-proton scattering cross section $\sigma_{\chi-p}$, the
neutralino-neutralino annihilation cross-section at zero temperature and
momentum $\left<\sigma_a v \right>$, and the neutralino relic density
$\Omega_\chi h^2$, all plotted versus $Z_g/Z_h = Z_g/(1-Z_g) $, the ratio of
gaugino and higgsino fractions (defined such that $Z_g + Z_h = 1$).  The range
of $\mu$ and $M_2$ used to produce Fig.~1 was 150 GeV $< \mu <$ 1.2 TeV and 8.9
TeV $> M_2 >$ 290 GeV. 

 From the plot we see that for mixed and gaugino-like
neutralinos ($Z_g > 0.1Z_h$) there is an excellent correspondence between
$\Omega_\chi h^2$ and $\left<\sigma_a v\right>$: as the annihilation cross
section in the Earth or Sun increases, the relic density decreases and
vice-versa, as expected.  Furthermore, we see evidence of the crossing symmetry
at work by examining the scattering and annihilation cross sections: as the
annihilation cross section increases, so does the scattering cross section.
This correspondence makes neutralino models with low relic densities but large
direct and indirect detection rates possible.  Fig.~1 shows that the crossing
symmetry is not effective when the neutralino has a small gaugino component
($Z_g < 0.1 Z_h$).  One reason the crossing symmetry is not perfect at low
$Z_g$ is because annihilation into W and Z pairs dominates for these
neutralinos. Hence, the annihilation cross section remains large, but the
scattering cross section decreases at low $Z_g$.  Moreover, at low $Z_g$, the
neutralino is almost a pure higgsino and is almost degenerate in mass with the
second lightest neutralino and the lightest chargino. Therefore,
coannihilations between neutralinos and charginos make $\left< \sigma_a v
\right>_{E.U.}$ larger than $\left< \sigma_a v \right>$. In fact, taking 
coannihilations out, we find
that the neutralino density without coannihilations, 
shown in Fig.~1 with the line labelled 
``w/o coanns", would be inversely proportional to $\left< \sigma_a v \right>$.

In the rest of this paper, we study the concrete case of the lightest
neutralino. For this purpose, we use a table of models allowed by all
accelerator limits (but without imposing any constraint from the recent
measurement of the muon anomalous magnetic moment \cite{g-2}). The table has
been produced with the DarkSUSY code \cite{darksusy} over the last few years
for other purposes, i.e.\ having in mind other issues which were addressed in
the papers of Ref.~\cite{bergstrom-gondolo,darksusypapers} for which the models
were originally computed.  Thus, to produce all the remaining figures, we have
not done any particular sampling of the models to favor lower densities and
higher detection rates.  We restricted our attention to models with
$\Omega_\chi \leq \Omega_{DM}= 0.3$ ($\Omega_\chi h^2 \leq 0.15$) for which we
found about 45,000 points in parameter space.

We use this table of models to show the validity of Eq. (2). In Fig.~2 we plot $ \Omega_\chi
h^2$ versus the zero momentum (and zero temperature) limit of the
neutralino-neutralino annihilation cross section times relative velocity
$ \left< \sigma_a v \right> $, together with the approximate relation 
in Eq. (2) (straight line). From the plot we see that Eq. (2) is a good approximation for the 
models with the largest  $ \left< \sigma_a v \right> $.
Generally, points are to the right and below the line because the cross
sections increase with energy and more annihilation channels, including
coannihilations, are open in the early universe. Some points are to the left
of the line due to resonances
active at low energy, or above the line due to coannihilations with particles
which have a lower annihilation cross section than neutralinos.

In a previous paper~\cite{dgg}, the same table of models was used to study the
possibility of direct detection of a subdominant component of neutralinos and
it was concluded that many models of subdominant neutralinos, even with WIMP
halo fractions as low as $10^{-4}$, are within the discovery limit of proposed
detectors in the conceivable future.

We now examine indirect detection of a subdominant neutralino component.

A proposed method of indirect detection consists of searching for gamma-rays
and rare cosmic rays produced by WIMP annihilations in dark
halos~\cite{indirect} or at the galactic center~\cite{gondolosilk}.  The rate
of these annihilations scales as
\begin{equation}
  \Gamma \sim \sigma_a \rho_\chi^2 .
\end{equation}
Since, as discussed above, $\rho_\chi$ scales as $\Omega_\chi$ and $\sigma_a$
scales approximately as $1/\Omega_\chi$, the annihilation rate in the halo
scales approximately as $\Omega_\chi$. As an example of halo fluxes, Fig.~3
shows the solar-modulated flux of anti-protons from neutralino annihilations in
the halo versus the neutralino relic density, together with the BESS 1-$\sigma$
measurement \cite{BESS} , both at 1 GeV kinetic energy and solar minimum.  The
envelope of maximum fluxes decreases linearly, not quadratically, with
decreasing density.  This shows that even if an increase in the cross section
compensates the decrease in one of the powers of the density, fluxes still
decrease linearly with the halo WIMP density.  Hence WIMPs which are a
subdominant component of the total CDM would be extremely difficult to detect
through this method, as already concluded in \cite{dgg,bottino2341}.  However,
we expect that the intensity of the high-energy neutrino emission from the Sun
and the Earth would in many cases remain high with decreasing density.

As the Sun and Earth orbit about the center of the galaxy, they sweep through
the dark matter halo.  Interactions with nuclei within the Sun and Earth slow
WIMPs enough such that the WIMPs can become gravitationally captured.  The WIMP
capture rate of a macroscopic body is given by \cite{gould}
\begin{equation}
  C = \pi \frac{\rho_\chi}{m_\chi} F\left(0\right) \sum_i
  \frac{(\sigma_{\chi-i})_s}{m_i} \beta_i M \left<v_{e,i}^2\right> 
  \left<v_{cut,i}^2\right>\sim
  \left<\sigma_s\right> \rho_\chi,
  \label{eq:2}
\end{equation}
where $\rho_\chi$ is the halo density of the WIMPs, $m_\chi$ is the mass of the
particular WIMP, $(\sigma_{\chi-i})_s$ is the elastic scattering cross section
of the WIMP with the $i$-th type of nucleus in the body, $\beta_i$ is the
fraction of the body's mass in that nucleus, $M$ is the total mass of the body,
$\left<v_{e,i}^2\right>$ is the squared escape velocity and
$\left<v_{cut,i}^2\right>$ is the squared speed at which capture is
kinematically cutoff, both averaged over the mass distribution of the $i$-th
element.  $F(0)$ is the zero-velocity phase space distribution of the WIMPs and
we call $ \left<\sigma_s\right>$ the averaged scattering cross section
appearing in the capture rate $C$.  Furthermore, the annihilation rate
$\Gamma_A$ of captured WIMPs within the body can be written in terms of the
capture rate as \cite{griest}
\begin{equation}
  \Gamma_A = \frac{C}{2} \tanh^2\left(\frac{t}{\tau_A}\right)
  \label{eq:3}
\end{equation}
where $t$ is the age of the body and $\tau_A = \left(C C_A\right)^{-1/2}$ is
the equilibration time between capture and annihilation.  $C_A$, the
annihilation rate per WIMP pair, can be written in terms of effective volumes
$V_1 $ and $V_2$ as $C_A = \left<\sigma v \right>_A V_2 / V_1^2$, where the
effective volumes depend on the pressure and temperature within the body in
question ($V_j = 3T_0/(2jm_\chi G\rho_0)$, where $T_0$ and $\rho_0$ are the core
temperature and density of the Sun or Earth).
  
When the equilibration time is much shorter than the age of the body, $\tau_A
\ll t$, the WIMP population is in equilibrium, i.e.\ the annihilation rate and
the capture rate are equal (apart from a factor of two).  In this case,
\begin{equation} 
  \Gamma_A \simeq \frac{C}{2}
  \sim \left<\sigma_s\right> \rho_\chi.
  \label{eq:100}
\end{equation}
If a decrease in WIMP relic density is compensated by a corresponding increase
in the scattering cross section, the annihilation signals from Sun and Earth
remain constant as the relic density decreases.

If instead the equilibration time is much longer than the age of the body,
$\tau_A \gg t$, then
\begin{equation}
  \Gamma_A \simeq \frac{C}{2}
  \left(\frac{t}{\tau_A}\right)^2 \simeq C_A C^2 t^2 \sim 
  \left<\sigma_a v\right>
  \left<\sigma_s\right>^2 \rho_\chi^2 \sim \left<\sigma_a v\right>
  \frac{\left<\sigma_s\right>^2}{\left<\sigma_a v\right>_{E.U.}^2} .
  \label{eq:101}
\end{equation}
Thus if the scattering cross section rises as the annihilation cross section
increases to decrease the current density of neutralinos, the annihilation rate
may remain high.  In this regime, overcompensation, namely a rise in the
neutrino flux despite decreasing relic neutralino densities, is possible.
Namely, if $ \left<\sigma_s\right> / \left<\sigma_a v\right>_{E.U.}$ is
constant, then $\Gamma_A \sim \left<\sigma_a v\right> \sim \left<\sigma_a
  v\right>_{E.U.} \sim \Omega_\chi^{-1}$. This overcompensation is different
from the increase in rates in direct and indirect detection qualitatively put
forward in Ref.~\cite{bottino2341} (see Fig.~2 in the first paper of
Ref.~\cite{bottino2341}), which for indirect detection would hold in the case
of equilibrium between capture and annihilation. Ref.~\cite{bottino2341} stated
that rates either remain constant or increase slightly with decreasing
densities.  Below we see that this statement holds only at relatively high
subdominant relic densities, while for smaller densities rates decrease (see
Fig.~5 below for indirect detection, and Fig.~2 in Ref.~\cite{dgg} for direct
detection).

Notice that $\tau_A=\left(C C_A\right)^{-1/2}$ is
\begin{equation}
  \tau_A \sim
  \left(\left<\sigma_s\right> \,
    \frac{\left<\sigma_a v\right>}
    {\left<\sigma_a v\right>_{E.U.}}\right)^{-1/2}.
  \label{eq:102}
\end{equation}
Thus, if both $\left<\sigma_a v\right>$ and $\left<\sigma_a v\right>_{E.U.}$
increase about equally, then $\tau_A \sim \left<\sigma_s\right>^{-1/2}$. Hence
a smaller $\tau_A$ corresponds to a larger scattering cross section. Since
capture is more efficient for large $\left<\sigma_s\right>$, the highest fluxes
will come from the models with the smallest $\tau_A$.

Figs.~4a and 4b illustrate this clearly\footnote{Notice that while in the other
  figures we use a regular grid of points covering the region with models, in
  Figs.~4, 7, 8 and 9c, d and e below we show the original points in the table
  of models. No meaning should be assigned to the density of points, as the
  latter is an artifact of the scanning used to generate the table of models
  (see discussion in Ref.~\cite{bergstrom-gondolo}).}.  There we plot the muon
rates in underwater or under-ice detectors (such as AMANDA \cite{AMANDA},
IceCube \cite{IceCube}, ANTARES \cite{ANTARES}, and NESTOR \cite{NESTOR}) from
annihilation in the Sun and the Earth as functions of the equilibration time
$\tau_A$.  (The vertical dashed line in figs.~3a and 3b denotes the age of the
solar system, $1.5 \times 10^{17}$ s.) These muons are produced by the neutrinos
generated in neutralino annihilations in the Earth or the Sun. These neutrinos
interact with matter in or near the underwater or under-ice detectors through
the charged current interaction $\nu N \rightarrow \mu X$ producing muons.  In
this paper, we take the energy threshold of detectable muons to be 25 GeV,
which is a reasonable value for IceCube, and assume that it will be possible to
detect a flux of 10 upward-going muons/km$^2$/yr.

For the Sun (Fig.~4a), $\tau_A$ can be $ \ll t$, $\simeq t$ or $ \gg t$. The
highest fluxes correspond to $\tau_A \ll t$, namely to models for which
equilibration between capture and annihilation has taken place. So for the Sun,
we do not expect overcompensation. In fact, in Fig.~5, which shows the
neutrino-induced up-going muon fluxes as a function of the relic density
$\Omega_\chi h^2$, we see compensation down to densities of $ \Omega_\chi h^2
\simeq 10^{-2}$, and a suppression of the rate for lower densities. In this
figure we include all neutralino masses, since the flux of neutrinos from the
Sun does not change much with neutralino mass. Present bounds from MACRO
\cite{MACRO}, Baksan \cite{baksan}, and Super-Kamiokande \cite{SK} exclude
fluxes higher than approximately 10$^3$ muons/km$^2$/yr.  Models with
$\Omega_\chi h^2 \simeq 1 \times 10^{-3}$ may have signals above 10
muons/km$^2$/yr. So a kilometer-size detector may be able to probe even
neutralinos which constitute only 0.7\% of the dark halo using neutrino signals
from the Sun.

The Earth is either in the regime where $\tau_A \gg t $ or in the regime in
which $\tau_A \simeq t$ (see Fig.~4b). Thus the highest fluxes correspond to
$\tau_A \simeq t$. Equilibration between capture and annihilation is not
reached, and so we expect cases with overcompensation. In fact, this is
apparent in Fig.~6a, where we plot the muon fluxes from neutralino
annihilations in the Earth as function of $\Omega_\chi h^2$ for neutralinos
lighter than 150 GeV. The envelope of the highest fluxes increases with
decreasing densities down to values of $\Omega_\chi h^2 \simeq 3 \times 10^{-3}$.
Close to this value of $\Omega_\chi h^2$, which corresponds to a neutralino
fraction in the halo of only 2\%, the muon fluxes become higher than 10
muons/km$^2$/yr, while signals from the Earth from neutralinos constituting the
whole of the halo would only reach the level of 1 event/km$^2$/yr (with a
threshold of 25 GeV). With the energy threshold of 25 GeV, neutralinos lighter
than 150 GeV are expected to be within reach of a kilometer-size neutrino
telescope searching for neutrino signals from the Earth only if they are
subdominant.

For neutralinos heavier than 150 GeV, the envelope of the highest muon fluxes
from the Earth remains almost constant with decreasing densities down to
$\Omega_\chi h^2 \simeq 5 \times10^{-3}$ (see Fig.~6b). There are expected rates
above 10 muons/km$^2$/yr for densities as low as $\Omega_\chi h^2 \simeq 3 \times
10^{-3}$, i.e.\ a kilometer-size neutrino telescope may detect neutrino signals
from the Earth for neutralinos with a halo fraction from 1 to as low as 2\%.

The predicted fluxes from the Earth are all below the current upper limits of
approximately $10^3$ muons/km$^2$/yr from MACRO \cite{MACRO}, Baksan
\cite{baksan}, Super-Kamiokande \cite{SK} and AMANDA \cite{amanda}. Notice that
in similar figures appearing in the literature (e.g. in Ref.~\cite{amanda})
there are models with signals higher than shown here. The reason is a lower
experimental energy threshold (e.g.\ 1 GeV) and a different choice of the
minimum density value the neutralinos can have to constitute the whole of the
halo ($ \Omega_\chi h^2 =$ 0.025 instead of 0.15).
 
If detection of a direct or indirect dark matter signal occurs, some model
discrimination could be achieved by looking for a signal in other ways.  This
is illustrated in Figs.~7a, 7b and 8a, 8b. 

In Fig.~7a we present the fluxes of up-going muons produced by neutrinos from
neutralino annihilations in the Earth versus relic density, for all models with
fluxes from the Sun larger than 10 muons/km$^2$/yr. This figure shows that only
some of the models detectable from the Sun are also visible from the Earth.
Contrariwise, all the models detectable from the Earth produce signals also
detectable from the Sun, as can be seen in Fig.~7b, where we present the fluxes
from the Sun of models with fluxes from the Earth above 10 muons/km$^2$/yr,
again versus relic density. Notice that the points plotted have densities
ranging from 100\% to 1\% of the halo dark matter density.  The big range of
fluxes in Fig.~7a is due to the fact that the capture rate in the Sun depends
on both the spin-independent and spin-dependent scattering cross sections,
while the capture in the Earth only depends on the spin-independent scattering
cross section.

The points in Fig.~7b visible from both the Earth and the Sun in kilometer-size
neutrino telescopes are also plotted in Fig.~8a, which shows that all of them
are within the sensitivity reach of the current generation of direct dark
matter detectors (such as CRESST II~\cite{cresst}, ZEPLIN II \cite{ZEPLIN},
CDMS-Soudan~\cite{CDMS-Soudan}, and EDELWEISS II~\cite{EDELWEISSII}).  Fig.~8a
presents the spin-independent scattering cross section of neutralinos with
protons multiplied by the neutralino halo fraction (the product that enters the
direct detection rate) versus the neutralino mass, for models producing fluxes
higher than 10 muons/km$^2$/yr from both the Earth and (therefore also) the
Sun.  In the figure we include the CDMS \cite{CDMS} and EDELWEISS 2000-2002
\cite{EDELWEISS} direct dark matter search limits, as these bounds are
currently the most stringent.  The CDMS and EDELWEISS 2000-2002 limits, which
we have not imposed on the models, eliminate several neutralino models with
high fluxes from both the Earth and the Sun.  However, many models fall outside
the current bounds but within the discovery limit of the current generation of
detectors.

Regarding indirect signals from the Sun, on the other hand, not even the larger
direct-search detectors of the next generation (such as ZEPLIN IV
\cite{ZEPLIN}, CryoArray \cite{cryoarray}, GENIUS \cite{GENIUS}, and XENON
\cite{XENON}) will be able to examine all models that give a detectable signal
from the Sun in a kilometer-size neutrino telescope. In fig.~7b, the models
producing more than 10 muons/km$^2$/yr from the Sun are plotted in the rescaled
spin-independent scattering cross section versus mass plane. Although several
models fall within the sensitivity reach of future direct detectors, several
others fall outside. The latter are those with small spin-independent cross
sections but large spin-dependent cross sections with protons in the Sun.

Independently of indirect signals, direct searches may reach much lower
densities.  Figs.~9a to 9e show all the models in our table separated in
decades of relic density $\Omega_\chi h^2$ or equivalently halo density fraction
$f_{CCDM}$.  In Fig.~9a neutralinos constitute between 100\% and 10\% of the
halo, in Fig.~9b between 10\% and 1\%, in Fig.~9c between 1\% and 0.1\%, in
Fig.~9d between $10^{-3}$ and $10^{-4}$ and in Fig.~9e between $10^{-4}$ and
$10^{-5}$ of the halo (the smallest density fraction being $2 \times 10^{-5}$).
In all these figures we plot the rescaled spin-independent scattering cross
section (i.e.\ the spin-independent scattering cross section of neutralinos
with protons multiplied by the neutralino halo fraction, which is the product
that enters the direct detection rate) versus mass. We also include the present
direct-detection bounds and future sensitivity limits.  We clearly see in these
figures that current data are probing neutralino halo fractions down to 1\%
(see Figs.~9a and 9b). In the next few years, data are expected from detectors
such as CRESST II~\cite{cresst}, ZEPLIN II \cite{ZEPLIN}, CDMS at the Soudan
mine~\cite{CDMS-Soudan}, and EDELWEISS II~\cite{EDELWEISS}. These experiments
will probe halo fractions as small as $10^{-4}$ (see Figs.~9a,b,c,d).  Further
in the future, another generation of detectors, such as ZEPLIN
IV~\cite{ZEPLIN}, CryoArray \cite{cryoarray}, GENIUS~\cite{GENIUS}, and
XENON~\cite{XENON}, may be able to reach down to neutralino halo fractions of
order $10^{-5}$, which are the smallest in our table of models.

In summary, indirect detection rates of neutralinos annihilating in the Sun and
the Earth remain above present detection thresholds even for neutralinos
constituting only 1\% of the halo dark matter. Indirect rates are drastically
reduced at lower density fractions. We find one instance in which the
detectability is highest around a halo fraction of 1\%, namely the case of
light neutralinos giving neutrino signals from the Earth. We do not find that
the detectability of relic neutralinos using signals from Sun or Earth is
usually favored for neutralinos of small $\Omega_\chi h^2$, as was previously
stated~\cite{bottino2341}.

Present limits from direct detection experiments probe down to 1\% of the halo
density, while data expected in the near future will probe down to $10^{-4}$
and the next generation of detectors may reach $10^{-5}$ of the halo density
(as already discussed in Ref. \cite{dgg}).

Many neutralino models constituting from 1\% to 100\% of the dark matter halo
are detectable in three ways (direct, indirect from the Sun, indirect from the
Earth), some in two, and some in only one.  Thus if a signal is found either in
direct detection experiments or in indirect searches for neutralino
annihilations in the Sun or Earth or in both, the question of which component
of dark matter was found, the primary or a sub-dominant one, may remain open.
However, if a signal would be seen from annihilation in the halo or in the
center of the galaxy, it would strongly point towards neutralinos being the
dominant component of the dark halo, since the signal from annihilation in the
halo decreases sharply with halo fraction.

\section*{Acknowledgments}

G.D.  and G.F.  were supported in part by the U.S.\ Department of Energy Grant
No.\ DE-FG03-91ER40662, Task C. This research was also supported in part by the
National Science Foundation under Grant No.\ PHY99-07949. J.E. thanks the Swedish Research Council for support.

We thank the Kavli Institute of Theoretical Physics at the University of
California, Santa Barbara, where this work was completed. We also thank Ted
Baltz, David Cline, Philippe di Stefano, Richard Gaitskell, Richard Schnee, and
Han-guo Wang for kindly providing data tables for some of the limit and
sensitivity curves of direct detection experiments.

\newpage

\section*{Figure Captions}

\begin{description}

\item[Fig.~1] Spin-independent neutralino-proton scattering cross section
  $\sigma_{\chi-p}$, zero momentum (and zero temperature) limit of the
  neutralino-neutralino annihilation cross section times relative velocity
  $\left<\sigma_a v\right>$, neutralino relic density $\Omega_\chi h^2$, and 
  $\Omega_\chi h^2$ if coannihilations would not occur (``w/o coanns"), as
  functions of the gaugino fraction $Z_g$ of the neutralino for a specific
  ``line'' of models with fixed neutralino mass and all other MSSM parameters
  held constant (see text).
  
\item[Fig.~2] Neutralino relic density $ \Omega_\chi h^2 $ versus the inverse
  of the zero momentum (and zero temperature) limit of the
  neutralino-neutralino annihilation cross section times relative velocity
  $\left<\sigma_a v\right>^{-1}$ together with the approximate relation in Eq.
  (2) (straight line). A regular grid of points shows the region covered with
  models.
 
\item[Fig.~3] Solar modulated anti-proton flux at 1 GeV kinetic energy at solar
  minimum (using the force-field approximation as implemented in DarkSUSY) as a
  function of the neutralino relic density.  A regular grid of points shows the
  region covered with models. The horizontal band shows the BESS 1-$\sigma$
  measurement of the antiproton flux.
  
\item[Fig.~4] Flux of muons in underground detectors with $E_\mu> 25$ GeV from
  neutrinos produced in neutralino annihilations in: (a) the Sun, (b) the
  Earth, as a function of the equilibration time in the Sun or Earth,
  respectively.  Each point represents an actual model.  The dashed line shows
  the current age of the Solar system, 1.5 $\times$ 10$^{17}$ seconds.
  
\item[Fig.~5] Flux of muons in underground detectors with $E_\mu> 25$ from
  neutrinos produced in neutralino annihilations in the Sun as a function of
  the neutralino relic density.  A regular grid of points shows the region
  covered with models. Approximate indirect detection current bounds and 
  future sensitivity of a kilometer-size detector are indicated.
  
\item[Fig.~6] As Fig.~5, but for neutralino annihilations in the Earth and
  neutralinos with: (a) masses $m_\chi \le$ 150 GeV, (b) masses $m_\chi >$ 150
  GeV.
  
\item[Fig.~7] Flux of muons in underground detectors from neutrinos from
  neutralino annihilations in: (a) the Earth, (b) the Sun, as a function of the
  neutralino relic density, for models which concurrently have muon fluxes from
  neutralino annihilations in (a) the Sun or (b) the Earth with rates greater
  than 10 muons/km$^2$/yr.  Each point represents an actual model.
  
\item[Fig.~8] Spin-independent neutralino-proton cross section times neutralino
  halo fraction as a function of neutralino mass, for models which (a) have
  muon fluxes from neutralino annihilations both in the Earth and in the Sun
  with rates greater than 10 muons/km$^2$/yr, and (b) have muon fluxes
  from the Sun with rates greater than 10 muons/km$^2$/yr.  Each point
  represents an actual model.  Best direct detection current bounds and
  selected future sensitivity limits are shown.
  
\item[Fig.~9] Spin-independent scattering cross section of neutralinos with
  protons times neutralino halo fraction versus mass, together with present
  direct detection bounds and future sensitivity limits. Models are separated
  in decades of halo fraction $f_{CCDM}$: (a) $0.1<f_{CCDM} \le 1$; (b)
  $10^{-2}<f_{CCDM} \le 0.1$; (c) $10^{-3}<f_{CCDM} \le 10^{-2}$; (d)
  $10^{-4}<f_{CCDM} \le 10^{-3}$; (e) $10^{-5}<f_{CCDM} \le 10^{-4}$ (the
  smallest value we found is $f_{CCDM} \simeq 2~10^{-5}$). In Figs.~9a and 9b,
  a regular grid of points show the region covered with models, in the others
  each point represent an actual model.

\end{description} 

\newpage
\includegraphics[width=0.89\textwidth]{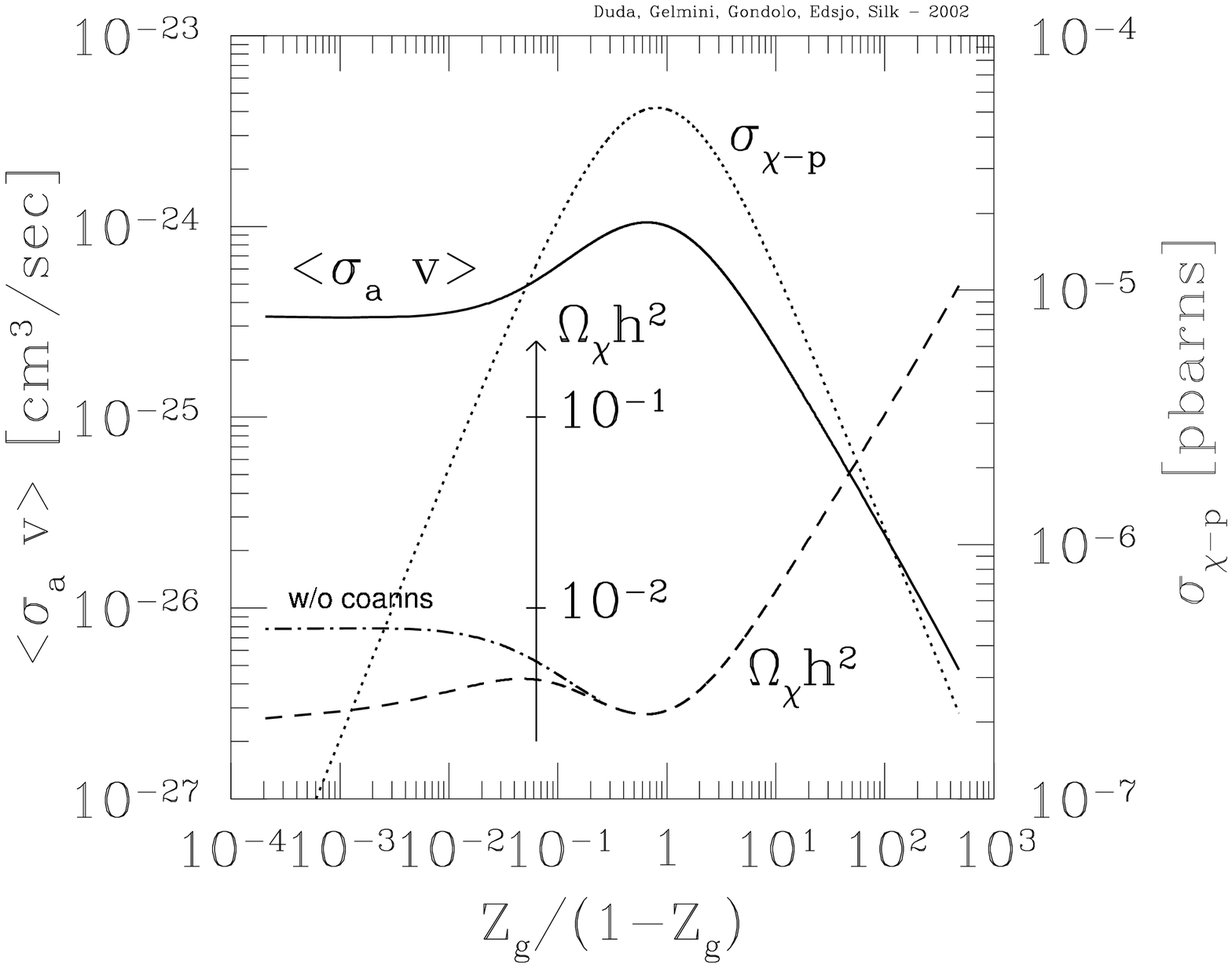}
\\Figure 1.
\newpage
\includegraphics[width=0.89\textwidth]{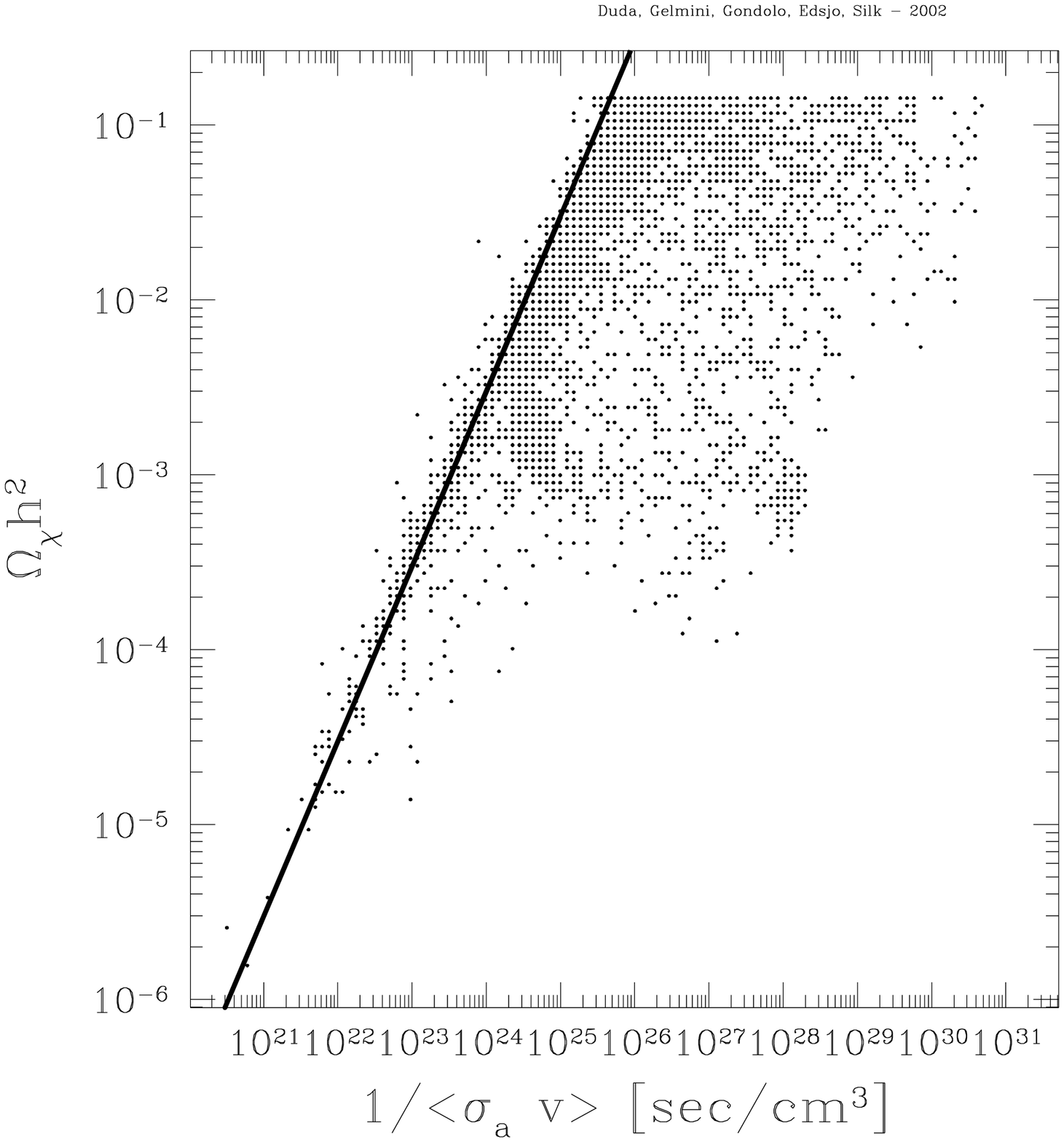}
\\Figure 2.
\newpage
\includegraphics[width=0.89\textwidth]{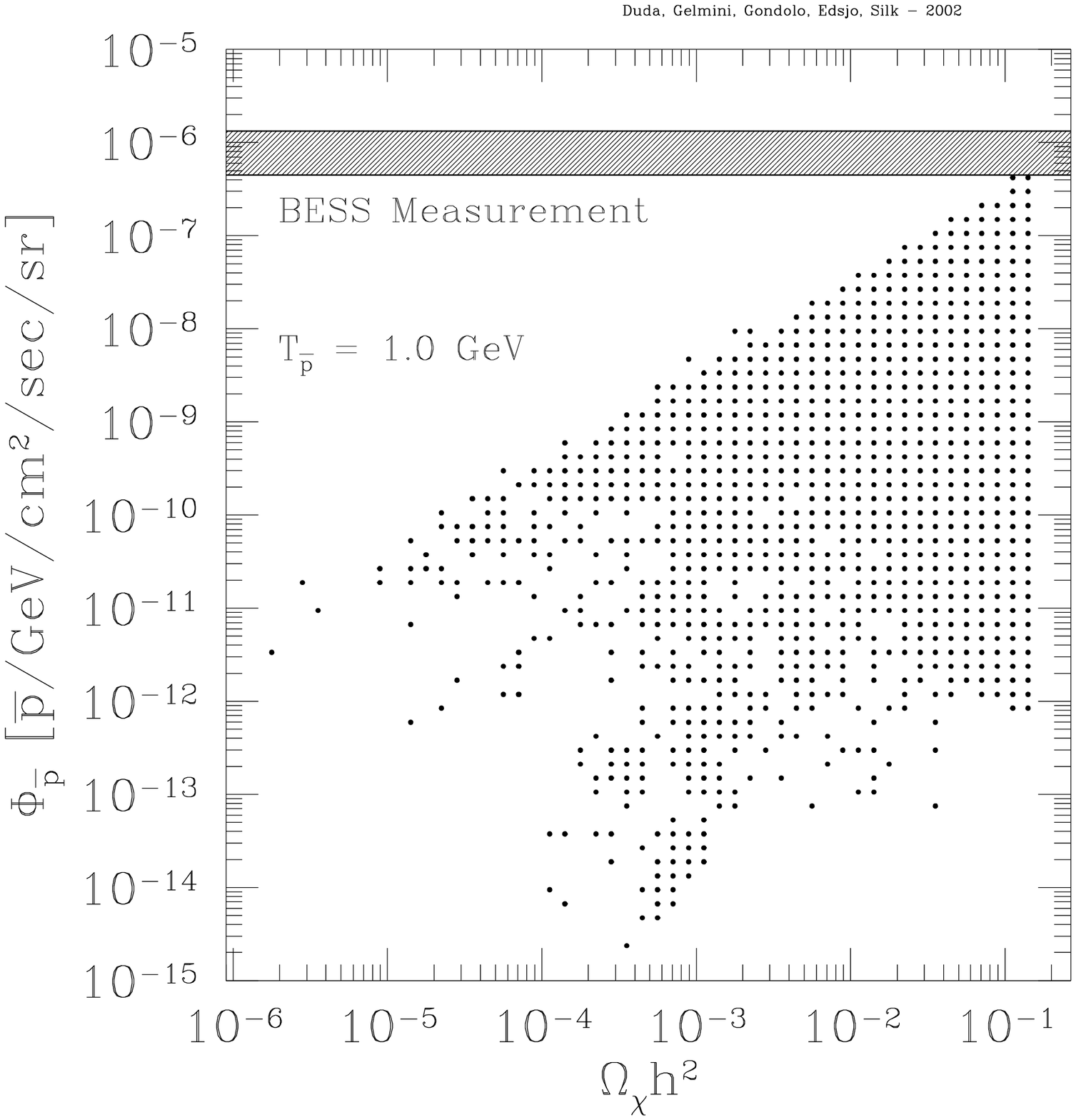}
\\Figure 3.
\newpage
\includegraphics[width=0.89\textwidth]{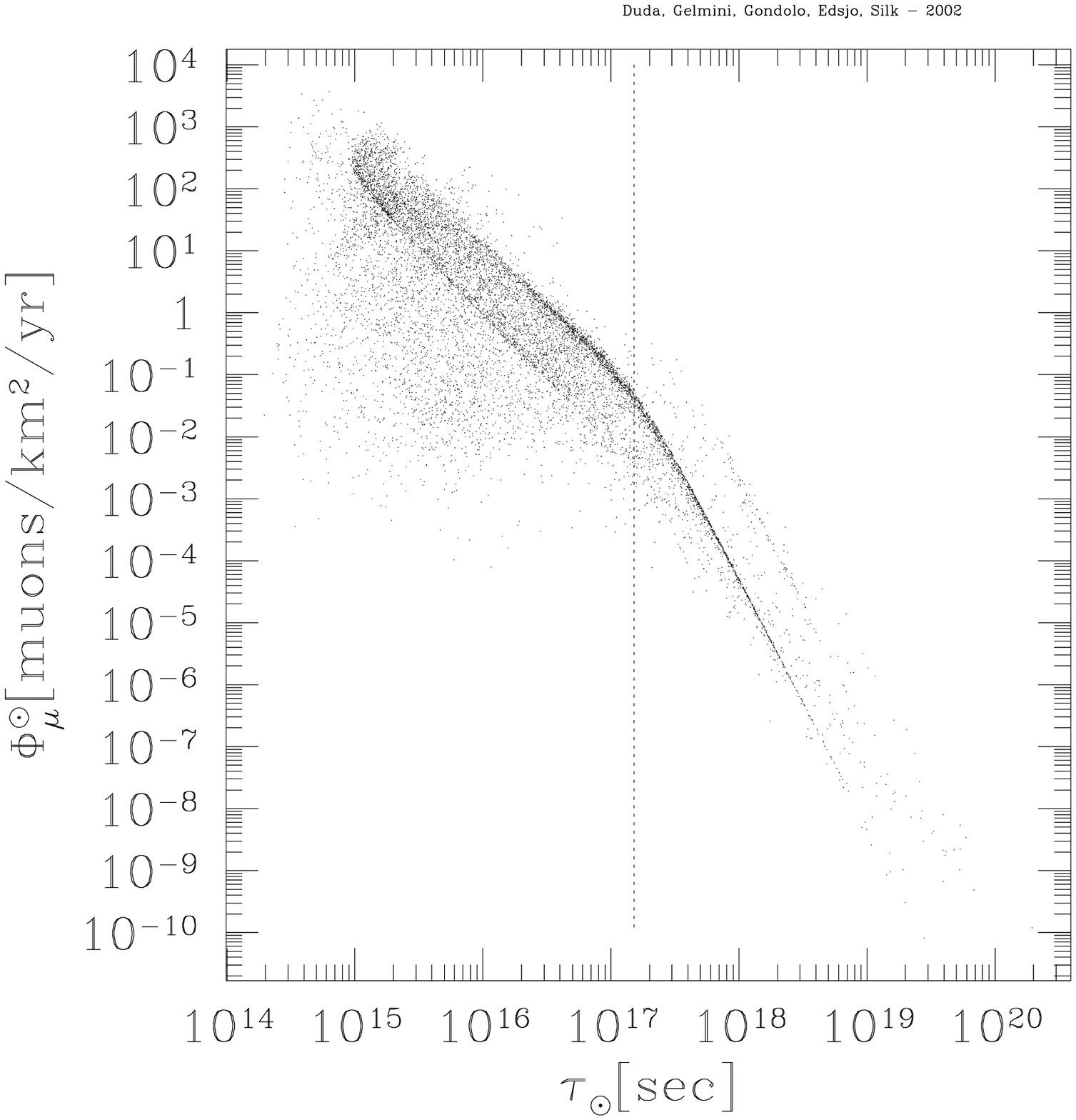}
\\Figure 4a.
\newpage
\includegraphics[width=0.89\textwidth]{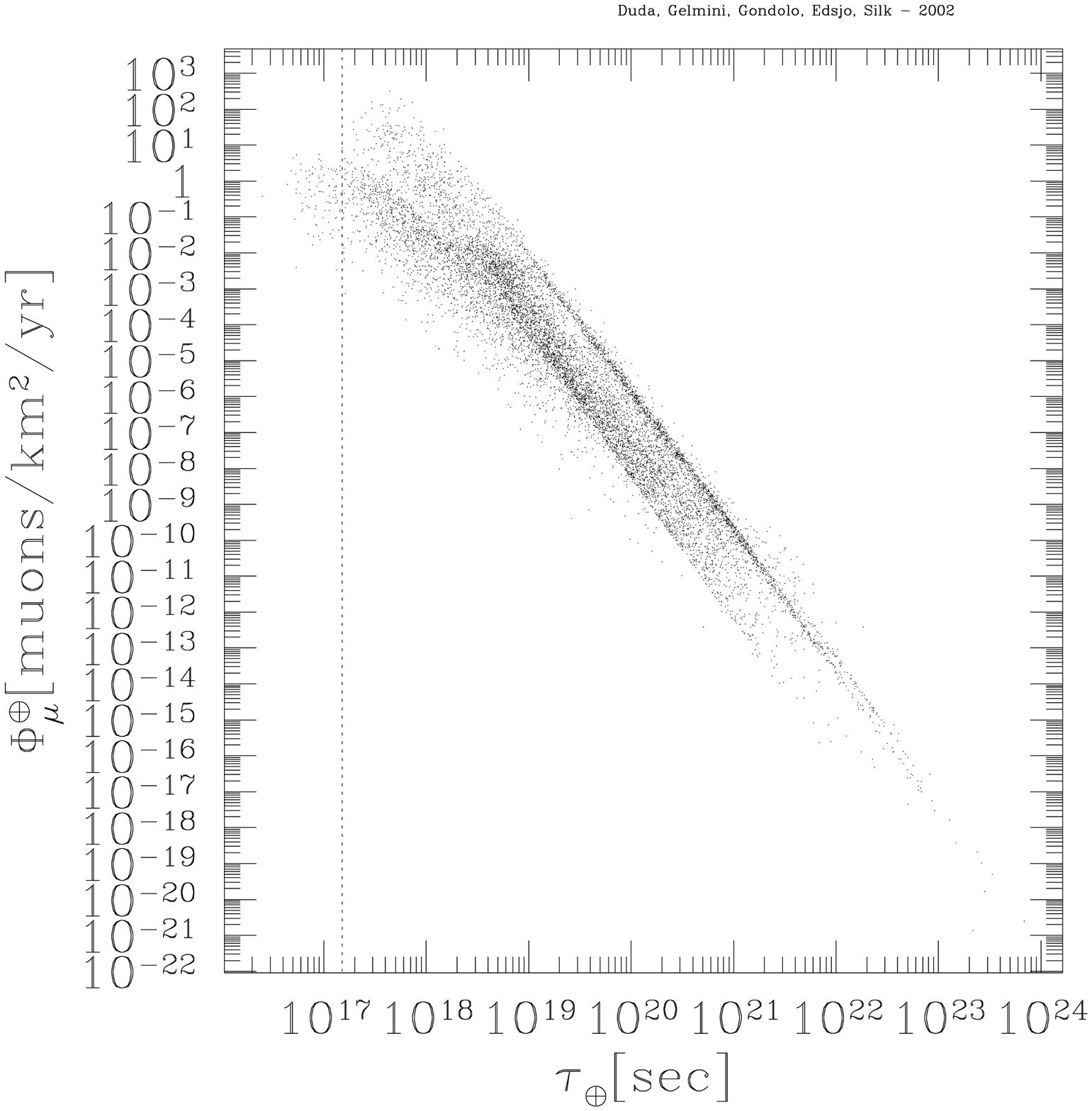}
\\Figure 4b.
\newpage
\includegraphics[width=0.89\textwidth]{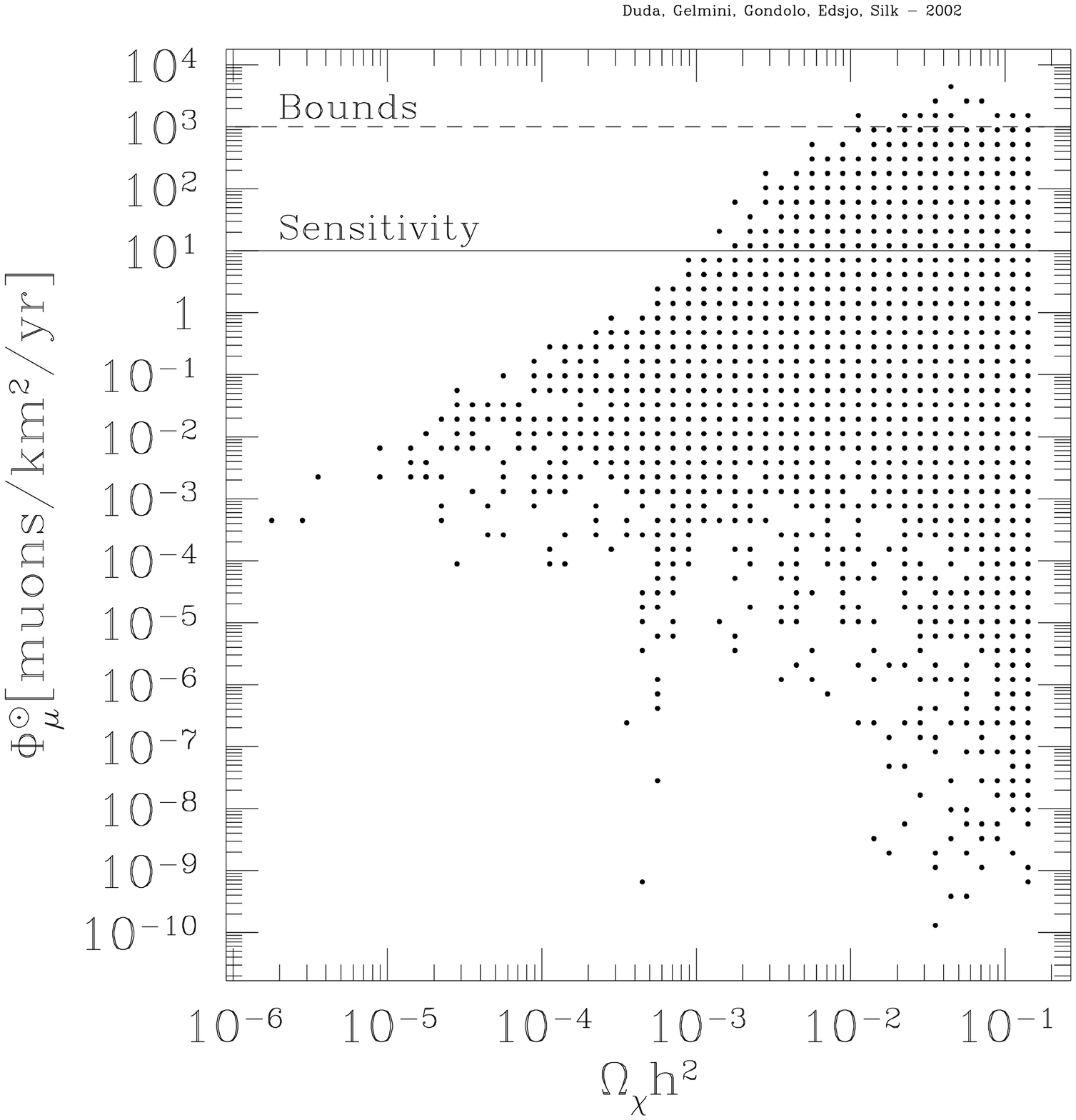}
\\Figure 5.
\newpage
\includegraphics[width=0.89\textwidth]{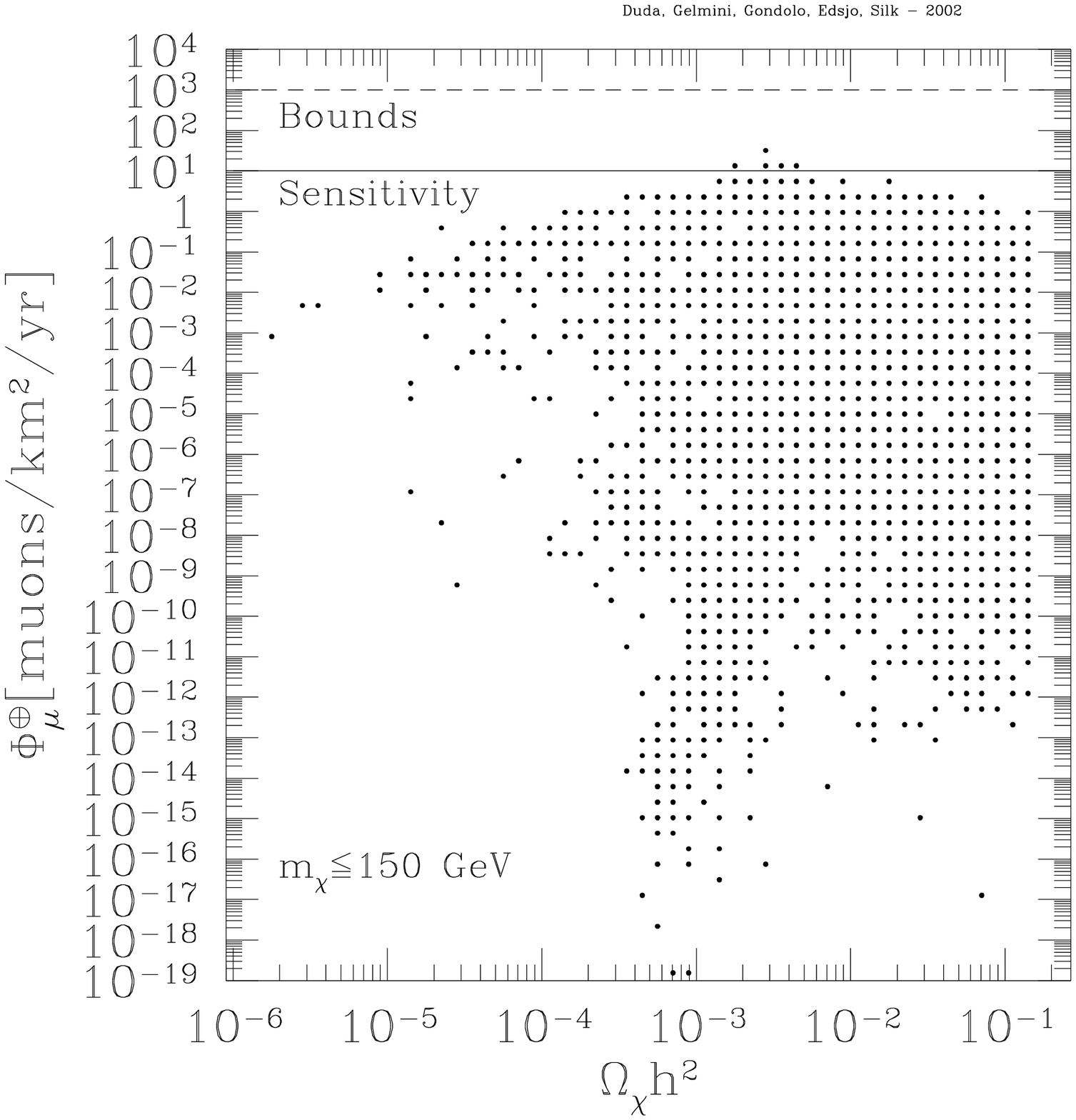}
\\Figure 6a.
\newpage
\includegraphics[width=0.89\textwidth]{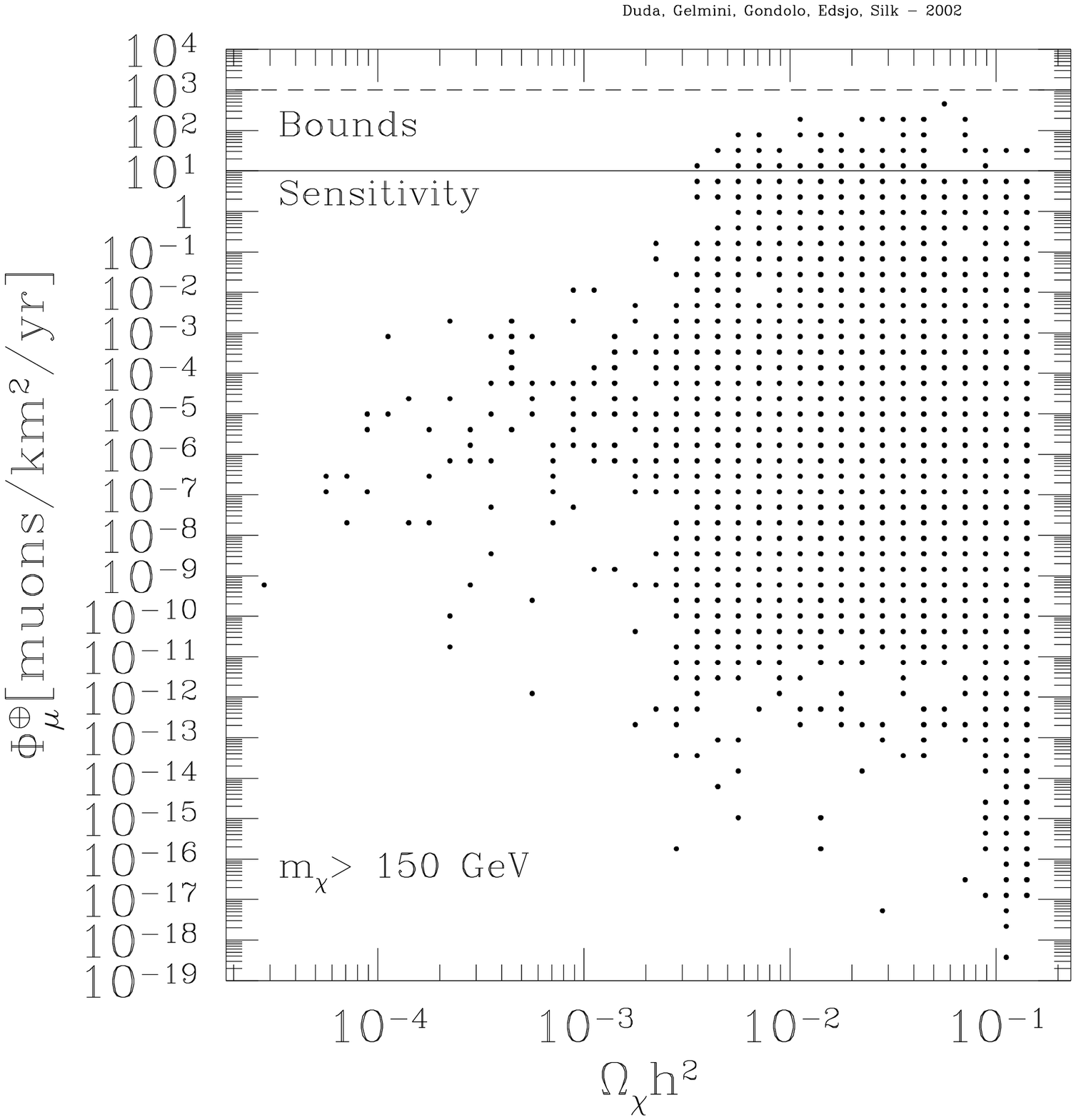}
\\Figure 6b.
\newpage
\includegraphics[width=0.89\textwidth]{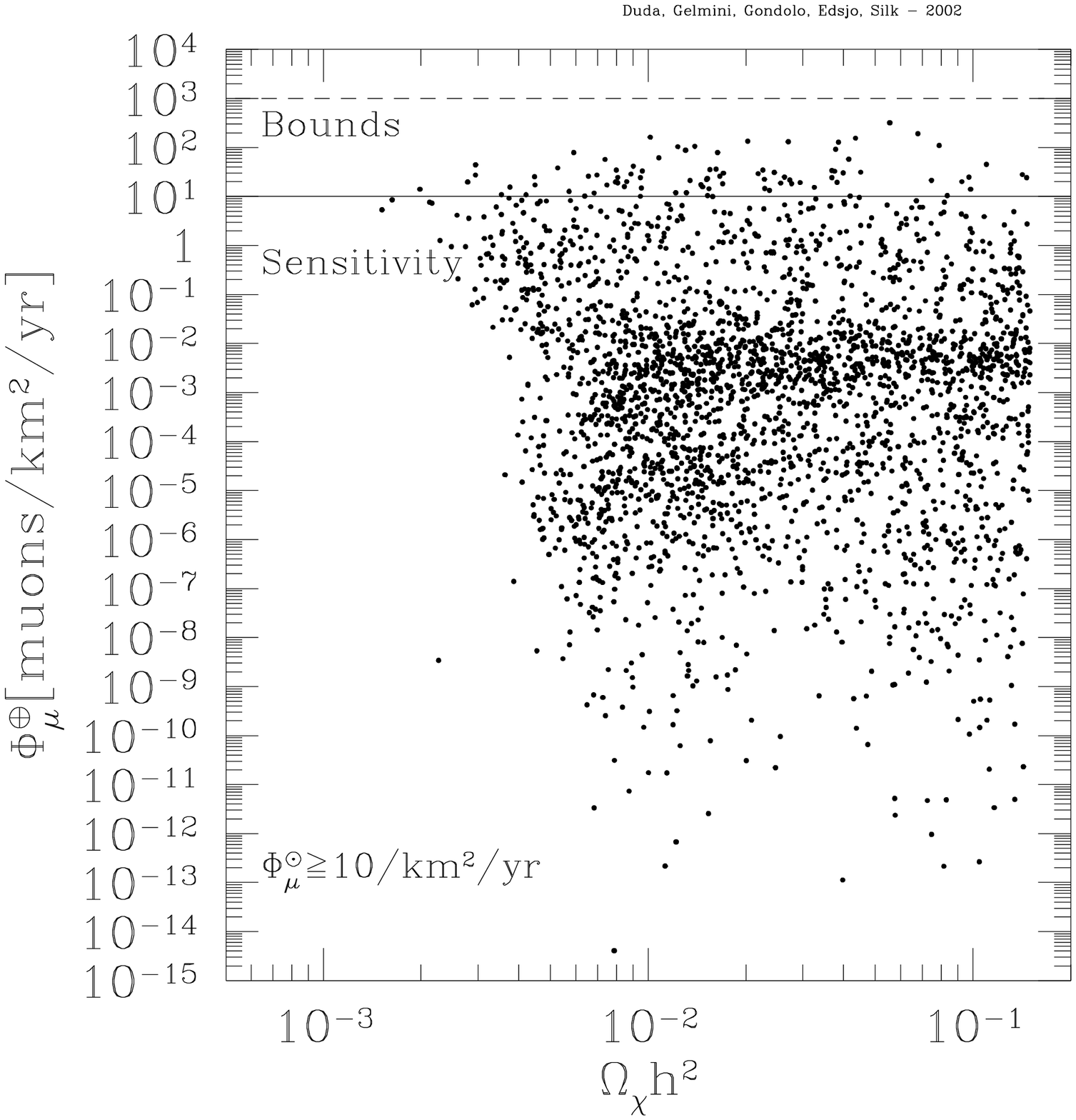}
\\Figure 7a.
\newpage
\includegraphics[width=0.89\textwidth]{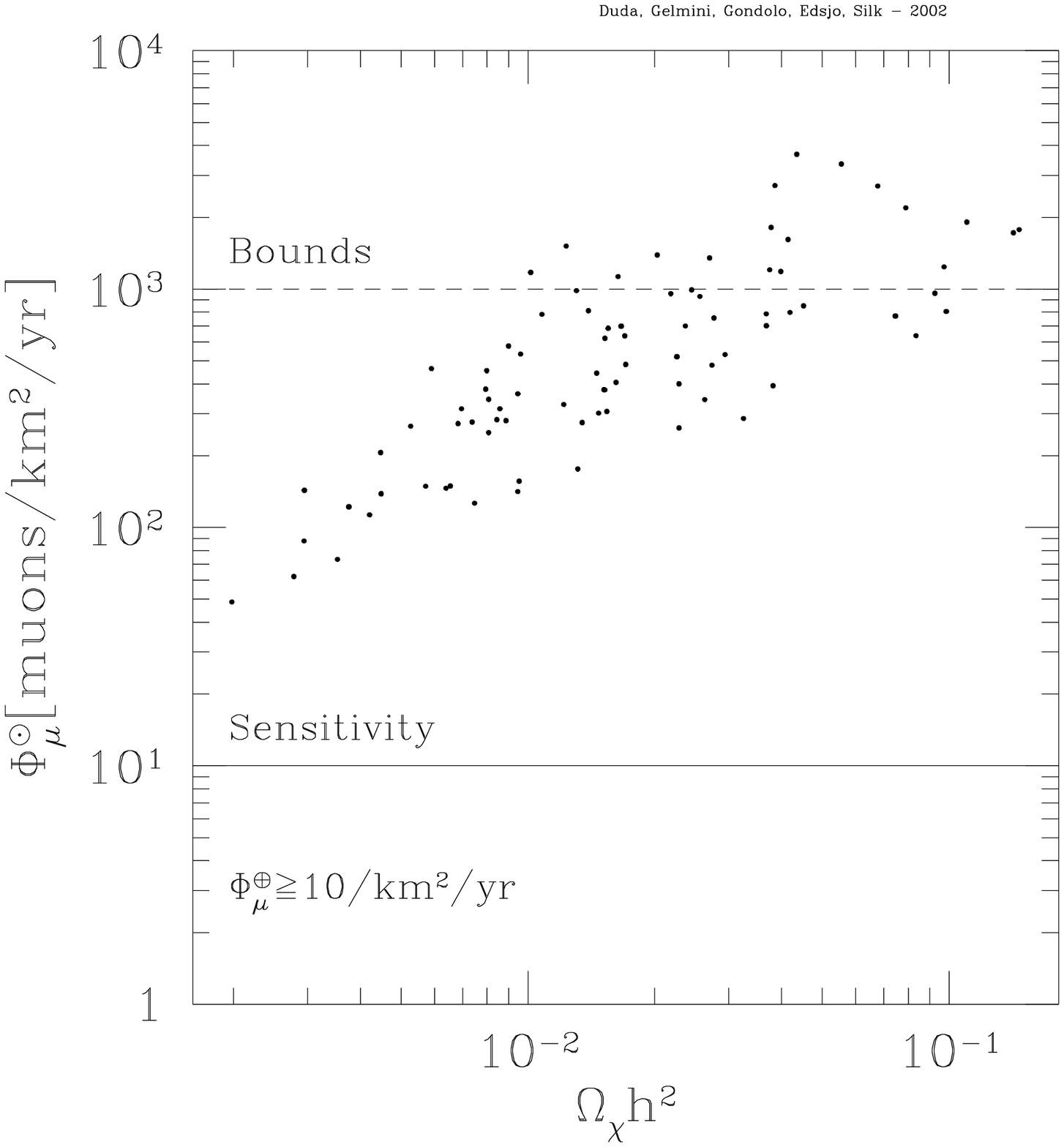}
\\Figure 7b.
\newpage
\includegraphics[width=0.89\textwidth]{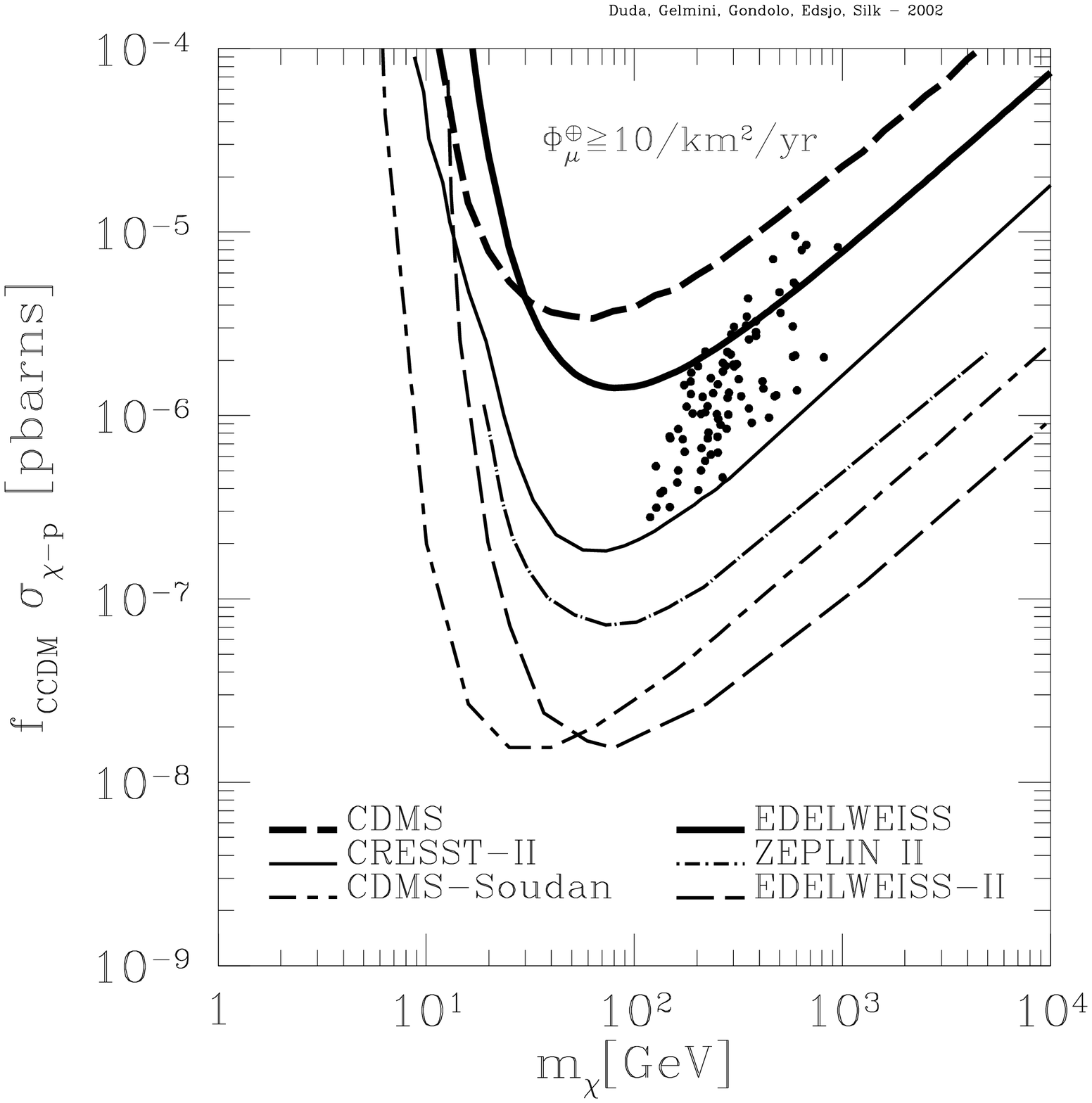}
\\Figure 8a.
\newpage
\includegraphics[width=0.89\textwidth]{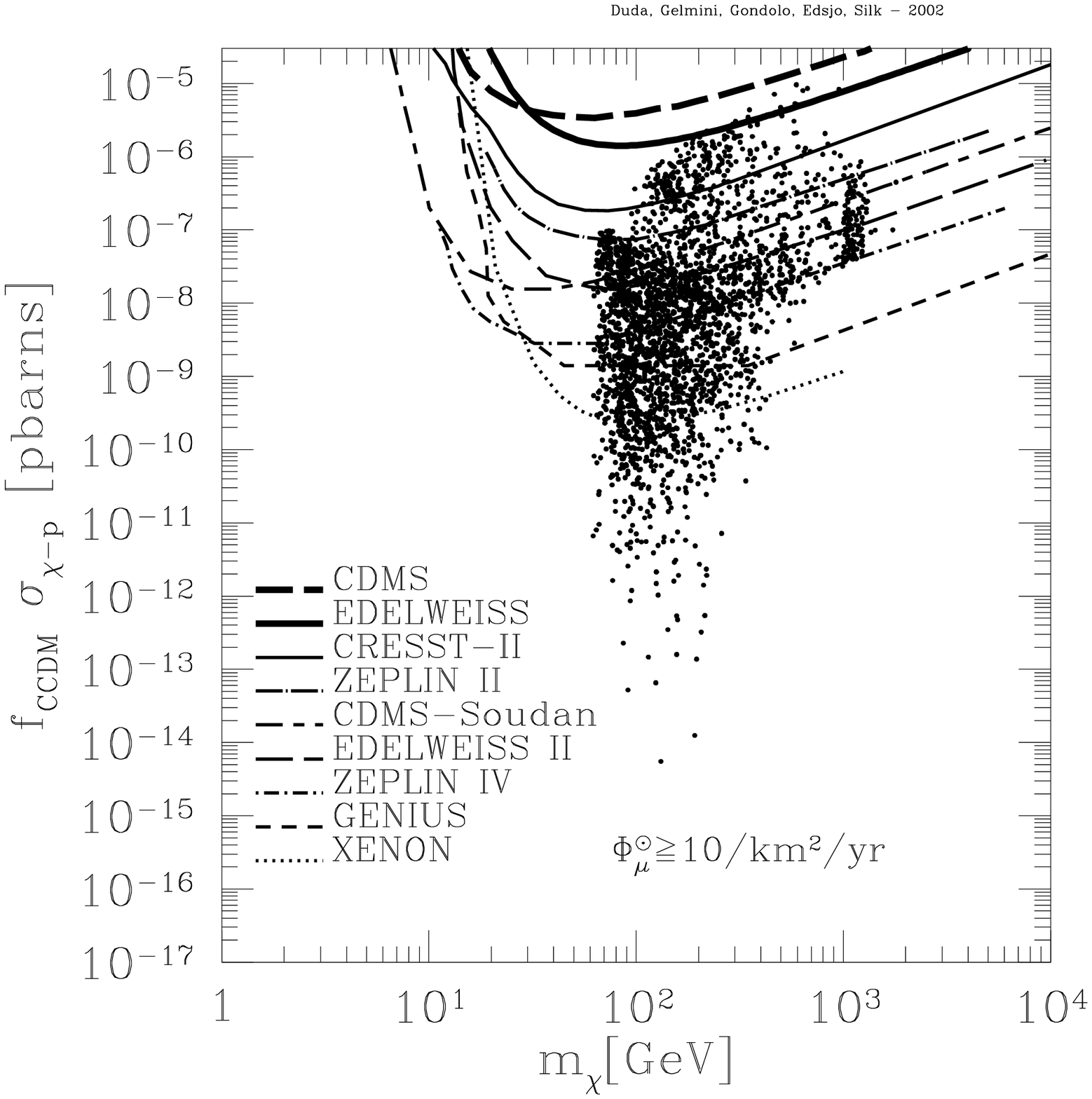}
\\Figure 8b.
\newpage
\includegraphics[width=0.89\textwidth]{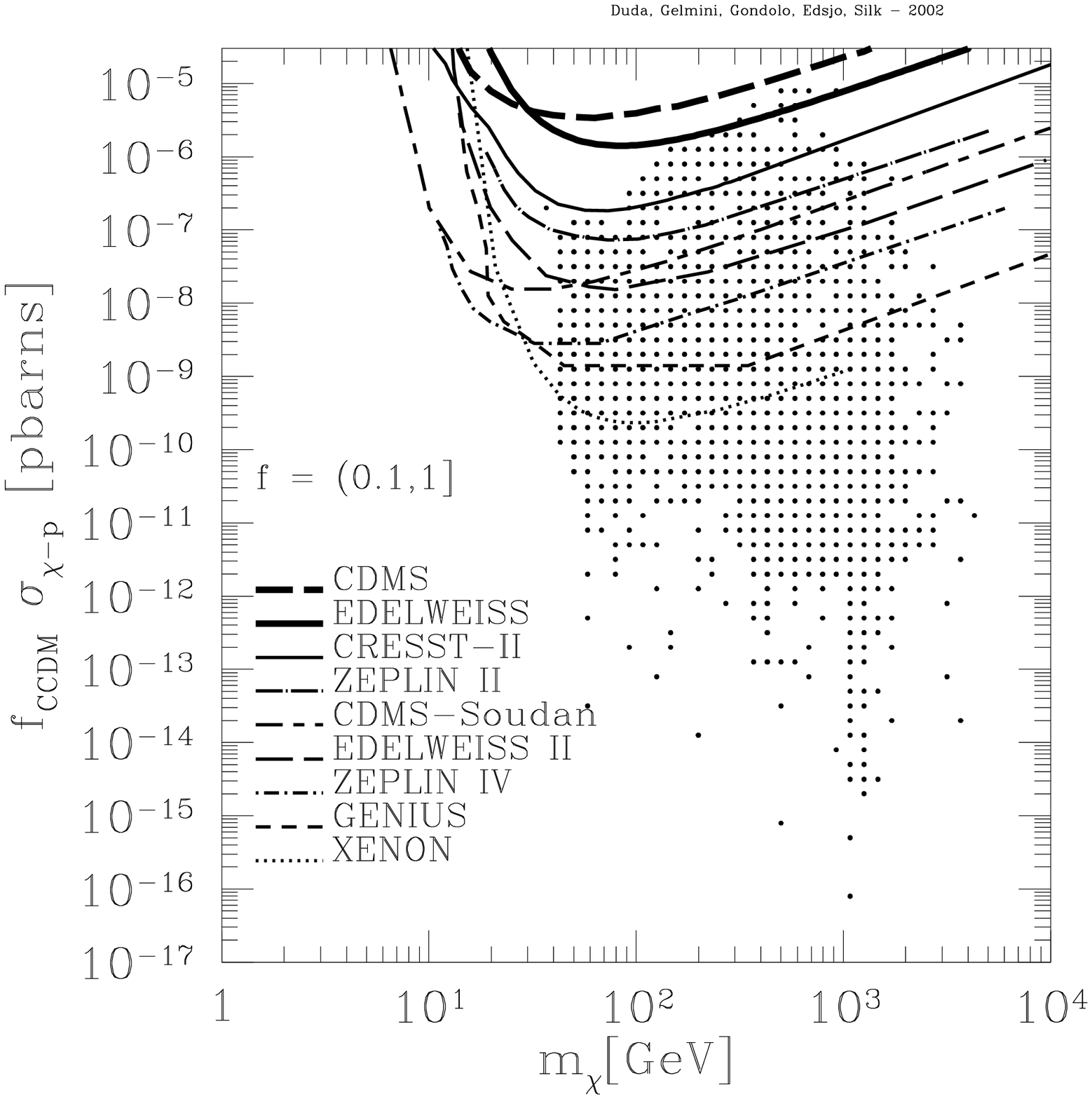}
\\Figure 9a.
\newpage
\includegraphics[width=0.89\textwidth]{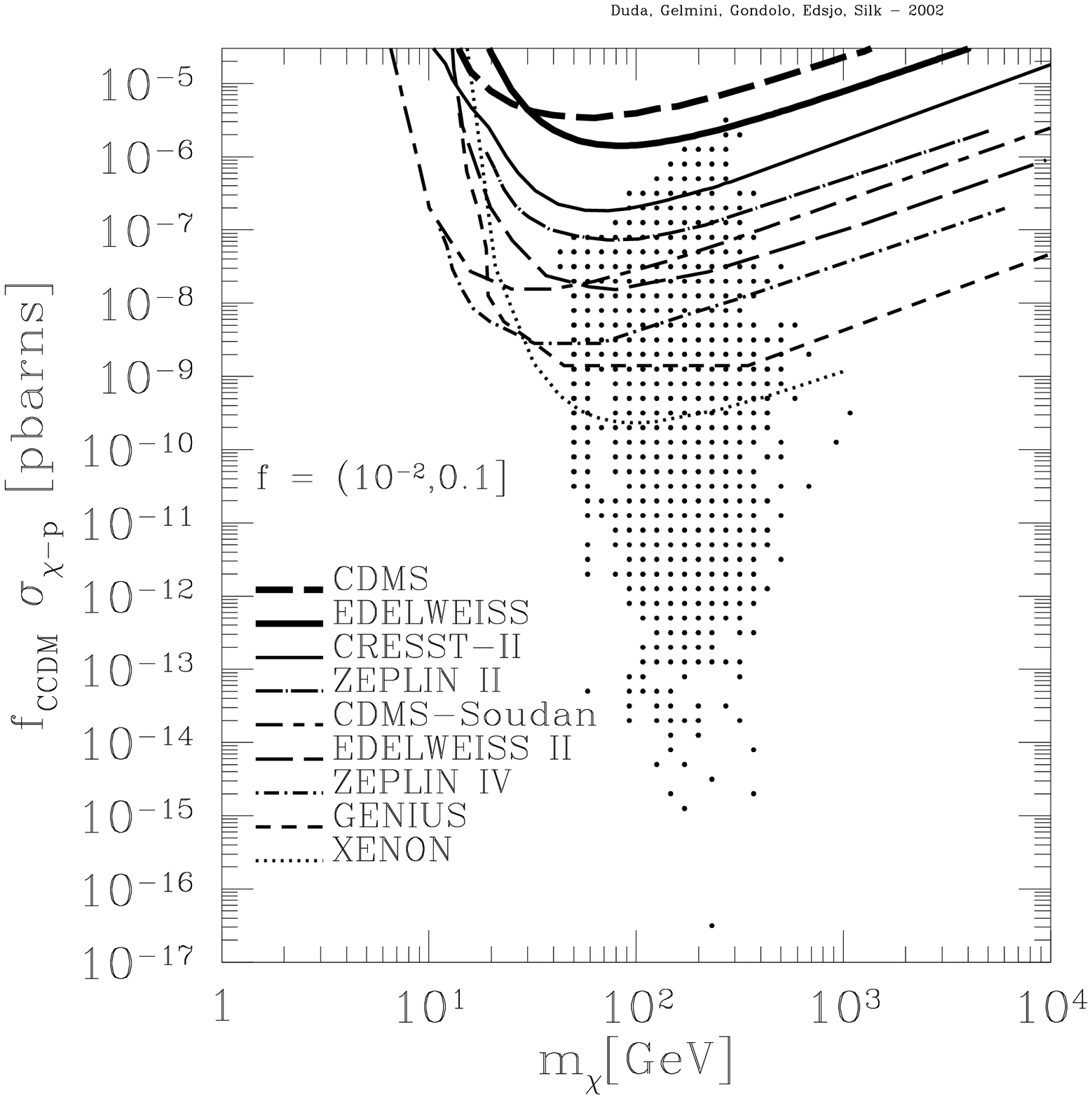}
\\Figure 9b.
\newpage
\includegraphics[width=0.89\textwidth]{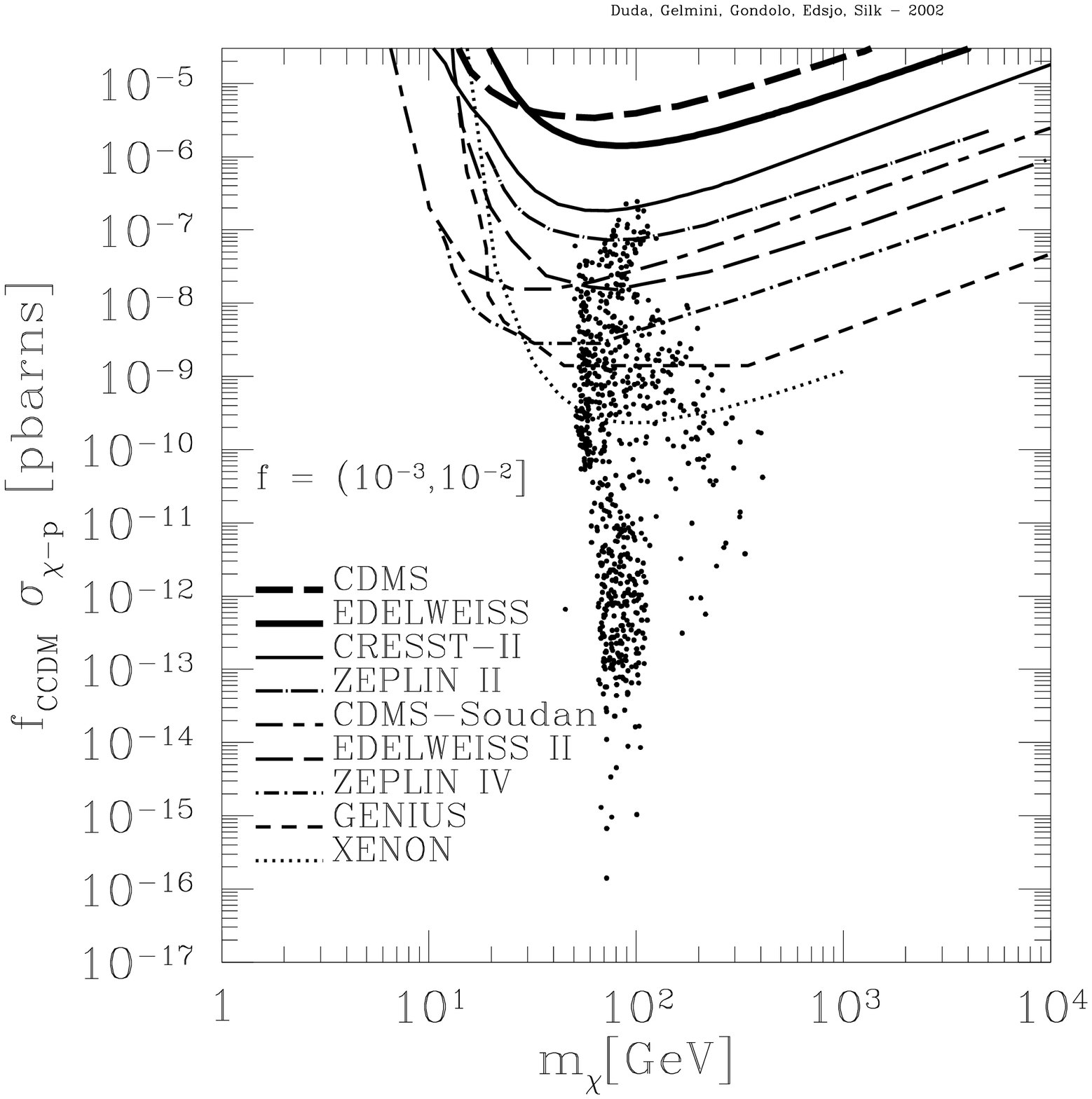}
\\Figure 9c.
\newpage
\includegraphics[width=0.89\textwidth]{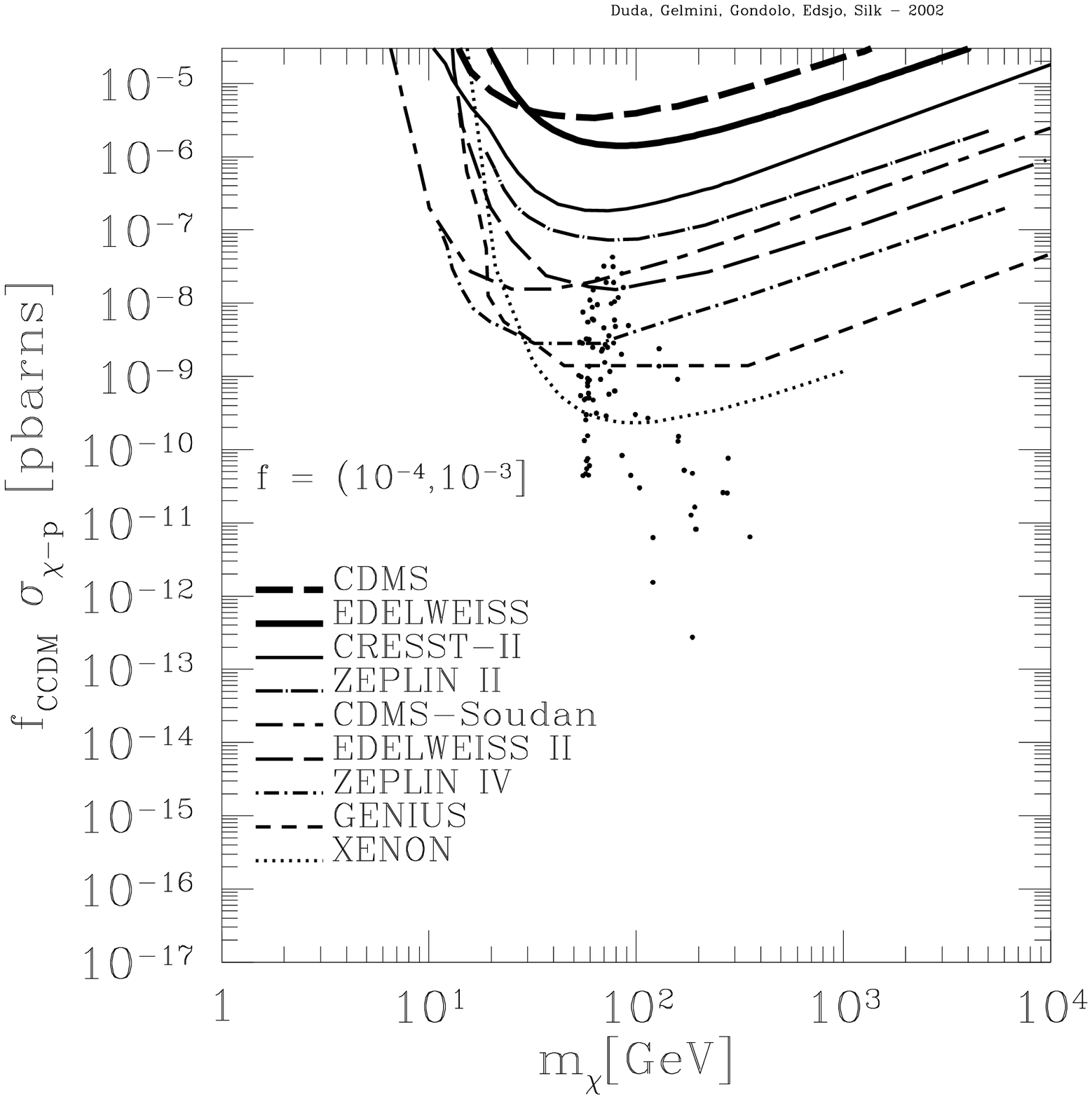}
\\Figure 9d.
\newpage
\includegraphics[width=0.89\textwidth]{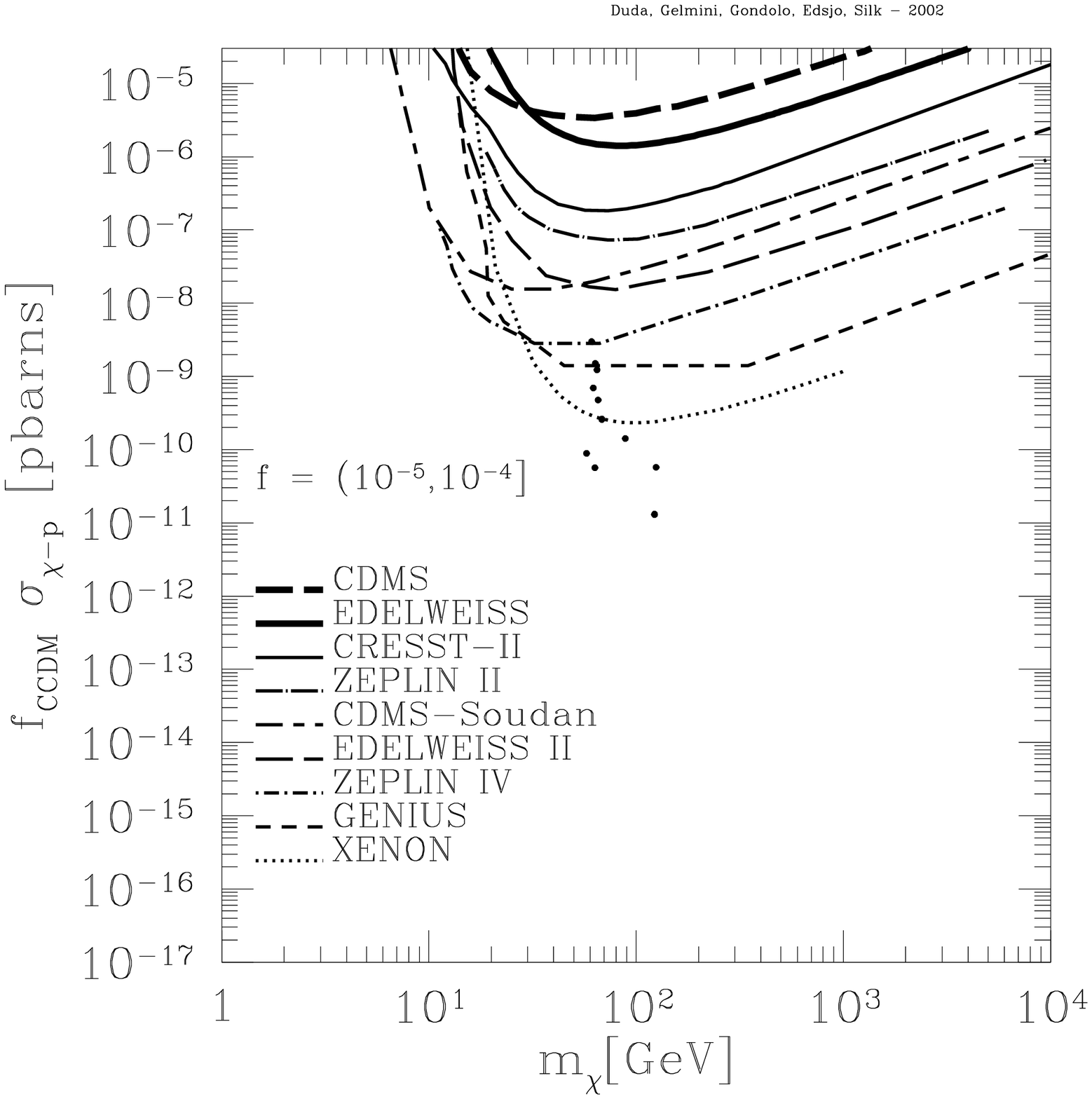}
\\Figure 9e.

\end{document}